\documentclass{article}
\usepackage{spconf,amsmath,graphicx}
\usepackage{color}
\usepackage{cite}
\usepackage{multirow}
\usepackage{booktabs}
\usepackage{url}
\usepackage{soul}

\title{THE CUHK-TUDELFT SYSTEM FOR THE SLT 2021 CHILDREN SPEECH RECOGNITION CHALLENGE}
\name{\begin{tabular}{c}Si-Ioi Ng$^{\dagger,1}$, Wei Liu$^{\dagger,1}$, Zhiyuan Peng$^1$, Siyuan Feng$^2$, Hing-Pang Huang$^1$,\\ Odette Scharenborg$^2$ and Tan Lee$^1$\end{tabular}
\thanks{$^{\dagger}$Equal contribution.}
}
\address{$^{1}$Department of Electronic Engineering, The Chinese University of Hong Kong, Hong Kong\\
$^{2}$Multimedia Computing Group, Delft University of Technology, Delft, The Netherlands\\
\begin{footnotesize}
\texttt{\{siioing,louislau\_1129\}@link.cuhk.edu.hk, tanlee@cuhk.edu.hk}
\end{footnotesize}
}    
\begin{document}
%
\maketitle
\begin{abstract}
This technical report describes our submission to the 2021 SLT Children Speech Recognition Challenge (CSRC) Track 1. 
Our approach combines the use of a joint CTC-attention end-to-end (E2E) speech recognition framework, transfer learning, data augmentation and development of various language models. Procedures of data pre-processing, the background and the course of system development are described. The analysis of the experiment results, as well as the comparison between the E2E and DNN-HMM hybrid system are discussed in detail. Our system achieved a character error rate (CER) of 20.1\% in our designated test set, and 23.6\% in the official evaluation set, which is placed at 10-th overall.


\end{abstract}
\section{Introduction}


Training a high-performance deep neural network (DNN) based automatic speech recognition (ASR) system often requires good amount of speech data paired with text transcriptions \cite{Pan2020}. Although the state-of-the-art performance of a DNN based ASR system has been shown comparable to human speech recognition \cite{saon2017english}, the underlying assumption is that there is no significant domain mismatch (i.e. channel, speaking style difference) between training and test data. 
Previous studies showed that the recognition accuracy could drop drastically for a mismatch caused by speech pathology 
\cite{liu2019acoutical,qin2018automatic}, background noise \cite{meng2017unsupervised} or speaker's attributes (e.g. age group \cite{DBLP:journals/taslp/PotamianosN03,Fainberg+2016}, gender \cite{DBLP:conf/interspeech/HsuZG17}).

The development of an ASR systems for child's speech has gained increasing research interest in recent years
\cite{Fainberg+2016,Ahmad2017,DBLP:conf/icassp/KathaniaSAA18,Wu2019,Shahnawazuddin2020}. 
The development of such ASR systems  remains a challenging problem. One major obstacle is insufficient amounts of child speech data due to the difficulties in data collection, which limits the development of a large-scale ASR system.
Another difficulty is related to properties of child speech. Compared to adult speech, child speech has higher fundamental frequencies (F0) due to shorter vocal tracts and smaller vocal folds \cite{das1998improvements}, slower and more variable speaking rates due to less  developed  articulators \cite{DBLP:conf/interspeech/PotamianosNL97},  and exhibits higher inter-speaker acoustical variability caused by the developing speech organs \cite{hermansky1990perceptual}. Moreover, children   tend to commit more pronunciation and grammatical mistakes due to the on-going language acquisition \cite{DBLP:journals/csl/ShivakumarG20}. 

To address the above issues, one popular approach is by exploiting strategies of data augmentation to increase the amount of child speech data, e.g. using the generative adversarial network (GAN) \cite{DBLP:conf/asru/ShengYQ19}, 
vowel stretch algorithm \cite{DBLP:conf/asru/NaganoFSK19}, voice conversion \cite{Shahnawazuddin2020}, or adding noise and reverberation \cite{Wu2019}.
Another research direction is based on transfer learning and adaptation. 
In \cite{Gale2019}, an acoustic model of child spech was first pre-trained with adult speech, which is followed by re-training with a limited amount of child speech data. The adversarial multi-task learning was used in \cite{Duan2020} to perform feature adaption between adult and child speech.  
The third strand focuses on the front-end feature engineering approaches. A spectral smoothing algorithm was proposed to mitigate the acoustic mismatch originating from the F0 difference \cite{Yadav2018}. In \cite{Shahnawazuddin+2016}, an adaptive-liftering based technique is utilized to derive pitch-robust features.

The encoder-decoder mode is commonly adopted in end-to-end ASR systems in recent years
\cite{Watanabe2018ESPnet}. While previous studies have shown that under some experimental settings the E2E architecture can outperform the hybrid DNN-HMM architecture \cite{li2020comparison}, not much has been investigated on deploying E2E systems in child speech recognition tasks.
In our submission, we adopt the joint CTC-attention learning framework to build an E2E ASR system for child speech. Transfer learning is performed by
incorporating adults' speech data to bootstrap an  ASR for child speech. Various speech data augmentation methods are utilized to mitigate over-fitting in the transfer learning based E2E system. 
N-gram and recurrent neural network (RNN) language models are developed to score the E2E ASR system in order to further improve the recognition performance.
In addition, a hybrid DNN-HMM based ASR system is constructed to compare the performance with the E2E system. 




\section{Data Preparation} 
\subsection{Description of CSRC dataset}
The dataset used in our system development is provided by the organizer of the Children Speech Recognition Challenge (CSRC). We follow the rules in Track 1\footnote{\url{https://www.data-baker.com/csrc_challenge.html}} of the CSRC, i.e., no external datasets to be used in system development.

Only the training data of the CSRC dataset is provided to the participants during  the system development period.
The training data consists of $3$ sets of Mandarin speech data, namely, Set A, C1 and C2.
Set A contains $341$ hours of adult read speech. Sets C1 and C2 include $29$ hours of child read speech and $30$ hours of child conversational speech respectively. 
Details of the training data 
are listed in Table \ref{tab:distribution}. 
All audio files are single-channel and recorded at 16 kHz sampling rate.

\vspace{-4mm}
\begin{table}[h!]
\centering
\caption{Details of the CSRC training dataset.}
\resizebox{0.9 \linewidth}{!}{%
\begin{tabular}{ c|c|c|c|c } 
\toprule
 Set& Hours &  No. of speakers & Age range & Style\\
\midrule
 A &341 & 1999 & 18-60 & Read \\
 C1 &29 &927 & 7-11 & Read  \\
 C2 &30  &54 & 4-11 & Conversational\\
\bottomrule
\end{tabular}
}
\label{tab:distribution}
\end{table}

\vspace{-4mm}
\subsection{Data pre-processing and partitioning}
\label{subsec:data_prep}
Data cleansing, generation of word pronunciation in the lexicon and data partitioning are carried out prior to ASR system development.
Upon examining the transcriptions of the CSRC dataset, 
Mandarin-English code-switching is found in some of the utterances. 
In our experiments, all English words are treated as out-of-vocabulary (OOV) words. 
The transcriptions contain inconsistent symbols to denote numbers, i.e.  both  Arabic and Chinese digits are used.  In our experiments, all Arabic digits are mapped to Mandarin digits. The punctuation marks in the transcriptions are also discarded.

A Mandarin pronunciation lexicon is not provided by the organizers. Two open-source toolkits are leveraged to 
generate the pronunciation lexicon, namely DaCiDian\footnote{\url{https://github.com/aishell-foundation/DaCiDian}} and Xpinyin\footnote{\url{https://pypi.org/project/xpinyin/}}. 
The created lexicon maps each Mandarin character into a sequence of  \textit{Initial} (onset) and   \textit{Final} (rime) tokens. The Initial is typically a consonant, while the Final contains a vowel nucleus followed by an optional consonant coda. 
Each vowel nucleus token has a tone label attached to it. The pronunciation lexicon is used in building the hybrid ASR system in our experiments.

The evaluation data in CSRC was not released to participants during the period of ASR system development. To facilitate system comparisons, 
we divided the provided training material into a  new training, validation and test sets with proportions of $81\%, 9\%$ and $10\%$ respectively.
The training-validation-test partitioning is done at speaker level for each of Sets A, C1, and C2.

\section{E2E ASR system description}
\begin{figure}[t]
\begin{center}
   \includegraphics[width=0.8\linewidth]{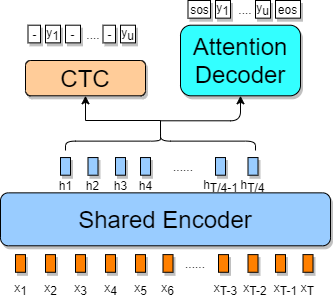}
\end{center}
\vspace{-6mm}
   \caption{Joint CTC-attention based end-to-end framework: 
   CTC is treated as an auxiliary task to help stable training and fast decoding of encoder-decoder architecture.}
\vspace{-3mm}
\label{fig:onecol}
\end{figure}

Our ASR system is based on an encoder-decoder model with joint the CTC-attention framework \cite{DBLP:journals/jstsp/WatanabeHKHH17},  
as illustrated in Figure \ref{fig:onecol}.
Given an acoustic feature sequence as the input, the shared encoder encodes it into a hidden vector sequence. The hidden sequence is fed into the connectionist temporal classification (CTC) 
block and Attention Decoder block. 
All blocks are jointly optimized using a  standard forward-backward neural network training strategy.

The connectionist temporal classification (CTC) structure aims to handle the mapping between the input and output sequences of variable length without explicit alignments. Given the input sequence $\mathbf{x}$, the probability of all possible output sequence $\Phi(\mathbf{y'})$ can be expressed as, 

\begin{equation}
    P(\mathbf{y}|\mathbf{x})=\sum_{\mathbf{\pi}\in \Phi(\mathbf{y'})} P(\mathbf{\pi}|\mathbf{x}),
\end{equation}
where $\mathbf{y'}$ is the prolonged version of $\mathbf{y}$ with repeated token and additional blank symbols. $\pi =(\pi_1, ..., \pi_T)$ denotes one possible token sequence, while
$P(\mathbf{\pi}|\mathbf{x})$ can be approximated as the product of the soft-max output over time $T$.
The training objective of CTC is maximizing $P(\mathbf{y}|\mathbf{x})$, where the loss function is defined as,
\begin{equation}
     \mathcal{L}_{CTC}= - \ln{P(\mathbf{y^*}|\mathbf{x})},
\end{equation}
where $\mathbf{y^*}$ represents the ground truth token sequence.

The attention based encoder-decoder structure aims to directly predict the output token sequence in an auto-regressive manner, without the alignment and the assumption of conditional independence.
The probability of the output sequence is denoted as,
\begin{equation}
    P(\mathbf{y}|\mathbf{x})
=\prod_{u} P(y_u|\mathbf{x},y_{1:u-1}),
\end{equation}
where $u$ is the time stamp, and $y_u$ is the output from the attention decoder at $u$. Typically two special tokens, "start-of-sentences (sos)" and "end-of-sentence (eos)", are inserted to the sequences to help decide the completion of the hypothesis. The loss function of the attention based encoder-decoder is given by, 
\begin{small}
\begin{equation}
    \mathcal{L}_{attention} = -\ln{P(\mathbf{y^*}|\mathbf{x})} = -\sum_{u}\ln{P(y_u^*|\mathbf{x},y_{1:u-1}^*)},
\end{equation}
\end{small}
where $y_{1;u-1}^*$ denotes the ground truth of the previous tokens.

To make full use of the advantages of both CTC and attention mechanisms, a multi-task learning (MTL) based loss function is derived as \cite{kim2017joint},
\begin{equation}
     \mathcal{L}_{MTL} = \lambda \mathcal{L}_{CTC} + (1-\lambda) \mathcal{L}_{Attention}.
\end{equation}
The task weight $\lambda$ is manually tuned between $0$ and $1$. 
The CTC objective function acts as an auxiliary task to help speed up the aligning process at both training and decoding stages, while the attention mechanism relieves the limitation of conditional independence assumed in CTC \cite{kim2017joint}. 

Both the shared encoder and the attention decoder utilize the self-attention based Transformer structure \cite{DBLP:conf/nips/VaswaniSPUJGKP17}, motivated by its superiority over other recurrent structures as shown in a recent study\cite{li2020comparison}. 


\section{Experiments}

\subsection{End-to-End ASR}
\label{subsec:exp_e2e}
The input to the E2E ASR system consists of 80 dimensional filterbank (F-bank) features  and 3 dimensional pitch features \cite{ghahremani2014pitch}. These features are extracted using Kaldi \cite{povey2011kaldi}.
The system is trained to predict Mandarin characters. 
It outputs $5596$ distinct units, including $5593$ Mandarin characters and $3$ special functional symbols that represent 
``unknown'', ``null'' and ``end of sentence'' 
respectively. 
The shared encoder of the E2E ASR system  consists of 12 \textit{TransformerEncoder} layers. 
The attention decoder has 6 \textit{ TransformerDecoder} layers. In between, the  self-attention  and the feed-forward sub-layers have $320$ and $2048$  hidden units, respectively.
The task weight $\lambda$ is empirically set as $0.3$. A dropout rate of $0.1$ is used to prevent over-fitting.
The E2E ASR system is first trained using  data in Set A, and then retrained  on the data of the  merged Sets C1 and C2. During re-training, all network parameters are updated. 
$3$-way speed perturbation is performed to augment the training  data in each Set. 

To evaluate the effectiveness of transfer learning  from adult speech (Set A) to child speech (Set C1 and C2),
another E2E ASR system is trained  only on the merged C1 and C2 set while keeping the other configurations unchanged.


\subsection{DNN-HMM Hybrid System}
A hybrid system is built to compare with the E2E system.
We first build a Gaussian mixture model - hidden Markov model (GMM-HMM) triphone acoustic model using Set A. 
The acoustic  features  consist of 13-dimensional  Mel-frequency  cepstral coefficients  (MFCC),  their  first-  and  second-order  derivatives,  and the Kaldi pitch feature which are extracted every 0.01 second.
Linear discriminant analysis (LDA), semi-tied covariance (STC) transform and feature space Maximum Likelihood Linear Regression (fMLLR) are applied to the input acoustic features \cite{duda2012pattern,gales1999semi,gales1998maximum}.

Next, we initialize the training of the DNN-HMM acoustic model using Set A, 
using the alignments generated by the trained GMM-HMM acoustic model. 
We adopt the factorized time delay neural network (TDNN-F) as the neural network architecture. It was shown effective in the recognition of English child speech \cite{Wu2019}. 
We use 40-dimensional hi-resolution MFCC + pitch feature as the input.
X-vectors are appended to each input as a strategy of speaker adaption \cite{snyder2018x}.

The TDNN-F system consists of 17 layers, with a hidden layer size of 1024 and a bottleneck size of 256. It is trained with combined objective functions of   cross-entropy and lattice-free maximum mutual information (LF-MMI). After training the TDNN-F, we perform the transfer learning by retraining the TDNN-F  with Sets C1 and C2. All layers are updated with a learning rate factor of 0.25. Same as in the E2E ASR training setting, 3-way speed perturbation is performed to augment training data in each Set. 



\vspace{-3mm}
\subsection{Resolving over-fitting} 
The over-fitting issue was   found in our E2E ASR model training during system development: the model resulted in an as low as $2\%$ CER on training data, and   highly unstable CERs on validation data at different training epochs. 
While 3-way speed perturbation is performed to the training data (see Section \ref{subsec:exp_e2e}), 
to further alleviate over-fitting, several other types of data augmentation methods were applied  on top of the 
speed perturbed child speech training data. These include spectral augmentation (known as SpecAugment) \cite{park2019specaugment}, reverberation augmentation using room impulse response (RIR) and volume perturbation \cite{ko2017study}. This resulted in an increase of the child speech training data to $450$ hours. The performance comparisons of both the E2E and hybrid ASR systems with and without applying the aforementioned additional data augmentation methods are studied in our experiments.

\vspace{-1mm}
\subsection{Language model training}\label{LM}
For the hybrid system, sequences of Chinese characters in the  transcriptions are split  into Mandarin words (consisting of one or multiple characters) with the Jieba toolkit \cite{jieba}. 
For the E2E system, word splitting is not performed. 
It is built based on individual Mandarin characters in order to limit the system  output dimension.


An n-gram LM is trained for each of Sets A, C1 and C2 via a maximum entropy approach \cite{rosenfeld1996maximum}. These LMs are further combined with linear interpolation and used to score the E2E system and the hybrid system.
For the hybrid system, a 5-gram word-level LM is built, whereas the E2E system uses a 4-gram character-level LM The training is implemented by the SRILM toolkit \cite{stolcke2002srilm}. 

Besides, long short-term memory (LSTM) based recurrent neural network LMs (RNN-LM) are trained.
They are used to compare against n-gram LMs in the scoring of the E2E and hybrid ASR systems. The LSTM consists of 2 hidden layers with a hidden layer size of 512. The training is implemented using  Kaldi. The RNN-LM used to score the E2E system is character based, while the one re-scoring the hybrid system is word based. 


\vspace{-2mm}
\section{Results and Analysis}

\subsection{Effect of transfer learning}\label{tl}
Table \ref{tab:comp_training_strategy} shows the comparison on character error rate (CER) results of the E2E ASR systems trained with (1) Set A only; (2) Set C1 \& C2 only; and (3) transfer learning approach, i.e. pre-training on Set A followed by Sets C1 \& C2. They are all evaluated on our designated test data selected from Set C1 and C2. 
It is shown that the model trained only on adult speech performs the worst. This is mainly due to significant domain mismatch between the adult and child speech. From Table \ref{tab:comp_training_strategy}, it is clearly observed that by using only in-domain child speech for training, although the data amount is only 17\% of adult speech, it results in a CER reduction of 11\% absolute.
The transfer learning based system   achieves the best performance, resulting in a further CER reduction of 4\% absolute compared to the system trained with child speech only. 



\begin{table}[t]
\centering
\resizebox{0.6 \linewidth}{!}{
\begin{tabular}{c|c|c}
 \toprule
Training strategy & Hour &  CER  \\
\midrule  
Adult   only & $828$  &  $38.5\%$ \\ 
\midrule
Child   only &  $144$ &  $27.6\%$ \\
Transfer learning &$828\rightarrow144$ &   $23.6\%$ \\
 \bottomrule
\end{tabular}
}
\caption{Comparison on E2E ASR systems trained with different strategies. CER results are evaluated on the combined test sets in C1 and C2. ``Hour'' denotes training hours.}
\vspace{-2mm}
\label{tab:comp_training_strategy}
\end{table}



We tested two other  methods in realizing transfer learning in our preliminary experiments: 1) Freeze all pre-trained layers. A linear input layer is added which serves as the adaptation for child speech;
2) only parameters of the top several layers of the attention decoder and the bottom several layers of the encoder are updated, inspired by   \cite{DBLP:journals/csl/ShivakumarG20}. 
We found the two above mentioned methods could result in unstable training and did not yield better performance than updating all layers, thus the experimental results are not reported in this paper. 
We hypothesize these two methods either limit the model capacity, or need careful tuning of learning rate. 
Considering the performance, we stick with the strategy of updating all layers in transfer learning.


\subsection{CER analysis at speaking style and speaker level }
We are interested to understand how E2E ASR systems trained with different strategies perform differently in: (1) read vs. conversational child speech; and (2) different child speakers.
For each training strategy mentioned in Section \ref{tl}, 
both the overall CER results and standard deviations of  per-speaker CER results on Sets C1 and C2 are reported in Table \ref{tab:spk_CER}, respectively.
The overall CER on C2 is consistently worse than on C1, as well as having larger variation of CER among each speaker. This is in line with the widely acknowledged fact that ASR on  conversational speech is more challenging than that on read speech, due to the high variability in conversational speech, e.g. hesitation, repetition, speaking rate, etc. 
Besides, the amount of read speech  training data (both adult and child) is around 12 times the amount of conversational speech training data, which  is also a reason leading to poorer performance on conversational speech.

Interestingly,  while the ``child only'' system performs better than ``adult only'' system on the combined test sets in C1 and C2 (see Table \ref{tab:comp_training_strategy}), ``child only'' under-performs ``adult only'' on the test set in C1. The observation indicates that with larger amount of adult read speech data for training, an E2E system could perform better on the child read speech recognition task, compared to an E2E system trained on limited amount of read and conversational child speech. 
To some extent, this reflects the data-hungry nature of E2E ASR systems. 
From Table \ref{tab:spk_CER}, it is also noted that
for both the ``child only'' and ``transfer learning'' systems, 
the per-speaker CER variation on C2 is smaller than that of the ``adult only'' system. This indicates the importance of including child  speech data in  E2E ASR system training to handle the conversational style  in test data.

\begin{table}[t]
\begin{center}
\resizebox{0.99 \linewidth}{!}{
\begin{tabular}{c|cc|cc}
 \toprule
& \multicolumn{2}{c|}{  CER on C1} & \multicolumn{2}{c}{CER on C2}  \\
Training strategy  & Overall & Per-speaker std. & Overall & Per-speaker std. \\
\midrule
Adult   only & $12.4\%$ &$ 4.8\%$ & $56.0\%$ & $ 11.2\% $ \\ 
\midrule

Child   only & $17.5\% $ &$ 5.7\% $& $34.3\%$ & $ 8.4\% $ \\
Transfer learning & $11.4\%$ & $ 4.2\% $ & $31.8\%$ & $ 8.8\% $ \\
 \bottomrule
\end{tabular}

}
\caption{CERs of different E2E ASR systems on read  (C1) and conversational (C2) speech. ``Overall'' represents overall CER results, ``Per-speaker std.'' represents  standard deviation values computed by per-speaker CER results. } 
\vspace{-5mm}
\label{tab:spk_CER}
\end{center}
\end{table}



\subsection{Effect of applying additional  data augmentation} 
\begin{table}[t]
\centering
\resizebox{0.85 \linewidth}{!}{
\begin{tabular}{c|cc|c}
 \toprule
Model & Data augmentation & LM &   CER  \\
\midrule  
\multirow{4}{*}{E2E}& speed & - & 23.6\% \\ 
 &  speed+others & - &  21.2\% \\
 &  speed+others & 4-gram (C) & 20.3\% \\
 &  speed+others & 4-gram+RNN (C) & 20.1\% \\

\midrule
\multirow{3}{*}{Hybrid} & speed & 5-gram (W) & 21.2\%  \\
 & speed+others& 5-gram (W) &20.6\% \\ 
 & speed+others& RNN (W) &20.1\% \\ 

 \bottomrule
\end{tabular}
}
\caption{CER results on transfer learning based E2E and  hybrid ASR systems with different data augmentation methods and different choices of LMs. CER results are evaluated on the combined test sets in C1 and C2. ``others'' denotes SpecAugment, RIR and volume perturbation.  ``C'' and ``W'' denote character and word LM respectively.}
\vspace{-3mm}
\label{tab:comp_data_aug}
\end{table}
As shown in Table \ref{tab:comp_data_aug}, by applying additional data augmentation methods (SpecAugment, RIR and volume perturbation) on top of 3-way speed perturbation,  the transfer learning based  E2E ASR system (without incorporating an external LM) achieves $21.2\%$ CER on the combined test sets of C1 and C2, outperforming the same E2E system but without the three additional data augmentation methods ($23.6\%$, reported both in the first row in Table \ref{tab:comp_data_aug} and in the last row in Table \ref{tab:comp_training_strategy}) by $2.4\%$ absolute. 


The effect of applying the additional data augmentation methods in the hybrid ASR system is also shown in 
Table \ref{tab:comp_data_aug}. 
 As can be seen, by fixing the LM as 5-gram,  the  additional data augmentation methods bring an additional $0.6\%$ absolute CER reduction to the hybrid ASR system 
 ($21.2\% \rightarrow 20.6\%$). 

\subsection{Fusion of different LMs}
\label{subsec:fusion_shallow}
Two types of LMs are used to score the E2E ASR system, namely a 4-gram  LM and an RNN-LM.
Experimental results are shown in Table \ref{tab:comp_data_aug}. As can be seen, the use of a 4-gram LM to score the E2E system brings 
$0.9\%$ absolute  CER reduction. 
By incorporating  an RNN-LM simultaneously with the 4-gram LM, a CER of $20.1\%$ is achieved, which is the best performance of the E2E ASR system.
We found in our experiments that it is crucial to balance the weight ratio between the 4-gram and RNN LMs, as well as the CTC and attention decoder weights. 
Otherwise, the recognition performance could degrade, comparing to the E2E system scored with only the 4-gram LM.
The use of the RNN-LM to rescore our hybrid system is also reported in Table \ref{tab:comp_data_aug}. The CER performance is comparable to the best performance achieved in the E2E system.

Eventually, we submit the best performing E2E to the Challenge organizer. Recognition on the official evaluation set reports a CER of 23.6\% and our submitted system ranks 10-th overall.



\vspace{-2mm}
\section{Conclusion}
This paper summarizes our development of ASR systems for the Children Speech Recognition Challenge.
A joint CTC-attention based E2E ASR framework is adopted. Various transfer learning strategies, data augmentation methods and different types of LMs are investigated. A hybrid DNN-HMM ASR system is built to compare against the E2E ASR system.
Experimental results show the effectiveness of data augmentation and transfer learning on both E2E and hybrid ASR systems. We find the recognition of child conversational speech is more challenging than child read speech. 
It is also concluded that using an external 4-gram LM to score the E2E ASR system is effective, while fusing additional LM (e.g. RNN-LM) requires careful tuning of parameters for the additional gain of performance. 
The best performance of the hybrid ASR system is still comparable to the best performance of the E2E ASR system.

\bibliographystyle{IEEEtran}
\begin{small}
\bibliography{strings,refs}
\end{small}
\end{document}